\documentclass[10pt]{article}
\usepackage{graphicx, amsmath, amssymb, bm}

\def\Title#1{\begin{center} {\Large #1 } \end{center}}
\def\Author#1{\begin{center}{ \sc #1} \end{center}}
\def\Address#1{\begin{center}{ \it #1} \end{center}}

\newcommand\pubblock{\rightline{\begin{tabular}{l} Proceedings of the Second Annual LHCP\\ \pubnumber\\
         \pubdate  \end{tabular}}}

\newenvironment{Abstract}{\begin{quotation} \begin{center} 
             \large ABSTRACT \end{center}\bigskip 
      \begin{center}\begin{large}}{\end{large}\end{center} \end{quotation}}

\newenvironment{Presented}{\begin{quotation} \begin{center} 
             PRESENTED AT\end{center}\bigskip 
      \begin{center}\begin{large}}{\end{large}\end{center} \end{quotation}}

\def\Acknowledgements{\bigskip  \bigskip \begin{center} \begin{large}
             \bf ACKNOWLEDGEMENTS \end{large}\end{center}}

\def\beq{\begin{equation}}
\def\eeq#1{\label{#1}\end{equation}}
\def\eeqn{\end{equation}}

\def\beqa{\begin{eqnarray}}
\def\eeqa#1{\label{#1}\end{eqnarray}}
\def\eeqan{\end{eqnarray}}

\let\bar=\overbar

\def\Dslash{\not{\hbox{\kern-4pt $D$}}}
\def\dslash{\not{\hbox{\kern-2pt $\del$}}}

\def\msb{{\bar{\ssstyle M \kern -1pt S}}}

\usepackage{color}

 \newcommand\pubnumber{ }

\newcommand\pubdate{\today}

\def\affiliation{
Physics Department, University of Florida, \\
Gainesville, FL 32611, USA}

\begin{document}

\large
\begin{titlepage}
\pubblock

\vfill
\Title{Model-Independent Searches Using Matrix Element Ranking}
\vfill

\Author{ Dipsikha Debnath }
\Author{ James S. Gainer }
\Author{ Konstantin T. Matchev }
\Address{\affiliation}
\vfill
\begin{Abstract}
Thus far the LHC experiments have yet to discover
beyond-the-standard-model physics.  This motivates efforts to search
for new physics in model independent ways.  In this spirit, we
describe procedures for using a variant of the Matrix Element Method
to search for new physics without regard to a specific signal
hypothesis.  To make the resulting variables more intuitive, we also
describe how these variables can be ``flattened'', which makes the
resulting distributions more visually meaningful.
\end{Abstract}
\vfill

\begin{Presented}
The Second Annual Conference\\
 on Large Hadron Collider Physics \\
Columbia University, New York, U.S.A \\ 
June 2-7, 2014
\end{Presented}
\vfill
\end{titlepage}
\def\thefootnote{\fnsymbol{footnote}}
\setcounter{footnote}{0}
%

\normalsize 


\section{Introduction}

Next year, the CERN Large Hadron Collider (LHC) will resume
operations, accessing energy scales which have never been explored in
collider experiments.  While there are many well-motivated ideas about
what new physics may be found at these new higher energies, one might
worry, based on the lack of convincing evidence for
new physics at the LHC thus far, that we are looking in the wrong
places.  Thus it is useful to consider methods that are as
model-independent as possible.  The ultimate goal would be analyses
that discover generic departures from the standard model (SM), without
\emph{any} reference to a signal hypothesis.

Multivariate analyses~\cite{MVA} (MVA) play an increasingly important
role in experiments.  One MVA, which has been used in the
four-lepton channel for the discovery of the Higgs Boson~\cite{Higgs Discovery}
and the measurement of its properties~\cite{Higgs Properties}, is the
so-called ``Matrix Element Method” (MEM)~\cite{MEM}, which utilizes
the likelihood for signal and background hypotheses, calculated
largely from theory.  Specifically, the variables used in such
analyses, such as {\tt MELA KD}~\cite{MELA} or {\tt MEKD}~\cite{MEKD}
involve the ratio of signal and background matrix elements. A natural
question, then, is whether we can perform useful analyses using only
the background (squared) matrix element.  Such analyses will not be optimized
for a specific signal hypothesis and will thus be relatively effective
in discovering arbitrary new physics.  We have addressed this
question~\cite{Debnath:2014eaa} and found that such analyses are
indeed possible and potentially very useful.

\section{Discovery from the Background Matrix Element}

\begin{figure}[th]
\begin{center}
\includegraphics[width=0.55\columnwidth]{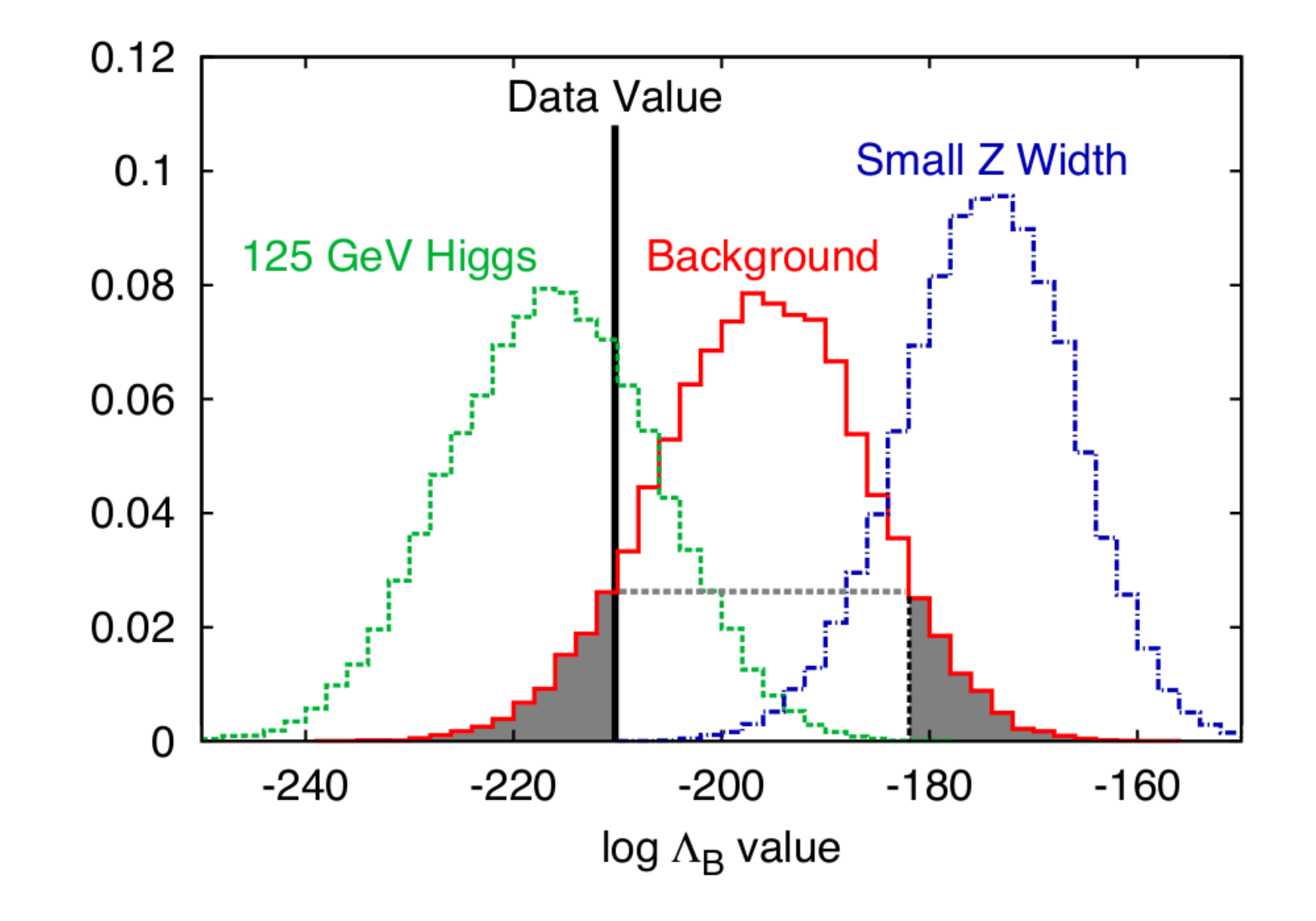} \\
\end{center}
\vspace{-15 pt}
\caption{\label{fig:log-mekd-pseudos}
The unit normalized distribution of $\Lambda_B$ for 20-event
pseudo-experiments for background $q\bar{q} \to 4\ell$ events (red solid
curve), $gg \to H \to 4\ell$ signal events for a $125$ GeV Higgs
(green dashed curve), and $q\bar{q} \to 4\ell$ events for which the
$Z$ boson width has been reduced by a factor of $5$ (blue dot-dashed
curve).  ``Data value'' refers to the value of $\Lambda_B$ for a particular 20 event
pseudo-experiment.
}
\end{figure}

As a test of the use of variables based on the squared background
matrix element, $|{\mathcal M}|^2_{bg}$,
we consider the sum of the natural logarithm of $|{\mathcal
  M}|^2_{bg}$, as calculated by {\tt MEKD}, a package for MEM
calculations for the four-lepton final state based on {\tt
  MadGraph}~\cite{MadGraph}.
In figure~\ref{fig:log-mekd-pseudos}
we plot the distribution this quantity, which we term $\Lambda_B$,
for pseudo-experiments with $20$ events from various
processes.  Clearly one can
distinguish various ``signal'' models from
the $q\bar{q} \to 4\ell$ ``background'', even using such a crude
analysis.  (More realistic analyses would use, e.g., a $\chi^2$ test
to determine how the observed distribution of $|{\mathcal M}|^2_{bg}$
matches the background distribution  of this quantity.)  It is also
worth noting that it is possible to have a signal hypothesis that is
``more background-looking" than the background itself, i.e., has a
distribution with support well to the right of the background
distribution in figure~\ref{fig:log-mekd-pseudos}.

\section{How to Flatten Background Distributions: Examples}

One could rule out the background hypothesis for describing the data using 
the method suggested by figure~\ref{fig:log-mekd-pseudos}.  However,
such an approach, which essentially uses only the mean of the quantity of
interest, is clearly not optimal.  A better approach would be to make
a binned histogram of $|{\mathcal M}|^2_{bg}$, as calculated for the
observed events, and use a $\chi^2$ technique to test the agreement
with the background distribution.

To do this, we must first determine where to set the bin boundaries.
A good choice is to choose these boundaries so that the expected
number of background events in each bin is equal.  To do this we can
``flatten'' the distribution, essentially by replacing $|{\mathcal
  M}|^2_{bg}$ with the cumulative distribution function (cdf) for this
quantity.  This is not a new insight, but we believe our main message,
that doing this allows for a more intuitive understanding of the
results of analyses such as those we are describing, is
under-appreciated.  In the hopes of encouraging the use of these
``flattened variables'' we describe several approaches to flattening.

\subsection{Flattening with Ranking}
\begin{figure}[th]
\begin{center}
\includegraphics[width=0.7\columnwidth]{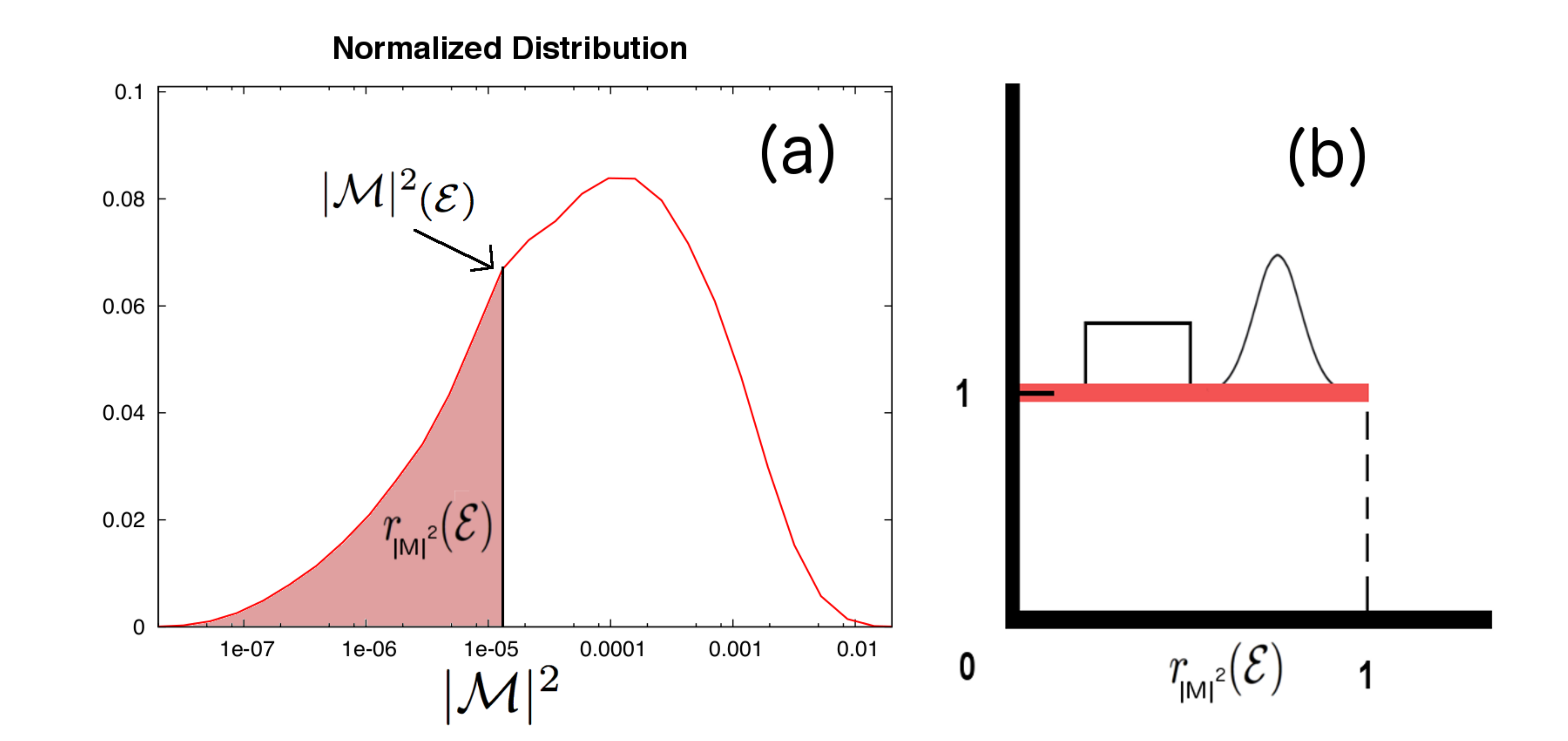}\\
\end{center}
\caption{\label{fig:ranking-cartoons}
In (a) we illustrate how a ``ranking variable'' is obtained from
distribution of background event {\tt MEKD} values, while (b) reminds
us that the background distribution of this ranking variable will be
flat, while signals will, in general, not be flat.}
\end{figure}

One approach is to take the normalized distribution,
$\frac{dN}{d\xi}$, for some kinematic variable, $\xi$,
and define a ``ranking'' variable,
\begin{equation}\label{eq: rank int}
r(\xi)=\int_{-\infty}^\xi \frac{dN}{d\xi^\prime} d\xi^\prime.
\end{equation}
which is the cdf for the background with respect $\xi$.  
The ranking, $r_\xi$, of an arbitrary event, ${\mathcal E}$,
is given by
\begin{equation}
r_\xi ( {\mathcal E}) = r (\xi( {\mathcal E})).
\end{equation}
In other words, the value of the ranking variable for the event, 
${\mathcal E}$, is equal to the 
fraction of background events which have values of $\xi$ less than the
value $\xi({\mathcal E})$ observed in the event ${\mathcal E}$.
This is illustrated in
figure~\ref{fig:ranking-cartoons}(a);
figure~\ref{fig:ranking-cartoons}(b) reminds us that the background
distribution of $|{\mathcal M}|^2_{bg}$ will be flat and that signals
will be identified by departures from flatness.  
While this ``flattening'' procedure works for any variable, $\xi$, 
the background matrix element will be a particularly sensitive choice
for particle physics events, as it contains information about the
resonance structure, spin correlations, etc. of the relevant background.


A useful property of $r(|{\mathcal M}|^2_{bg})$ is that it is easily
calculated from Monte Carlo (MC) events, as one can simply obtain the
value of $|{\mathcal M}|^2_{bg}$ for all of the events in a sample
after detector simulation, etc.  The fraction of events in the
resulting list, $\{|{\mathcal M}|^2_{bg, i}\}$, which have $|{\mathcal
  M}|^2_{bg}$ values less than a particular value, $|{\mathcal
  M}|^2_{bg,0}$ gives a good
approximation of $r(|{\mathcal M}|^2_{bg,0})$, that takes into
account all experimental effects that are well-modeled at the MC
level.

\subsection{Flattening with Quantile Bins}
\begin{figure}[th]
\begin{center}
\includegraphics[height=0.27 \columnwidth]{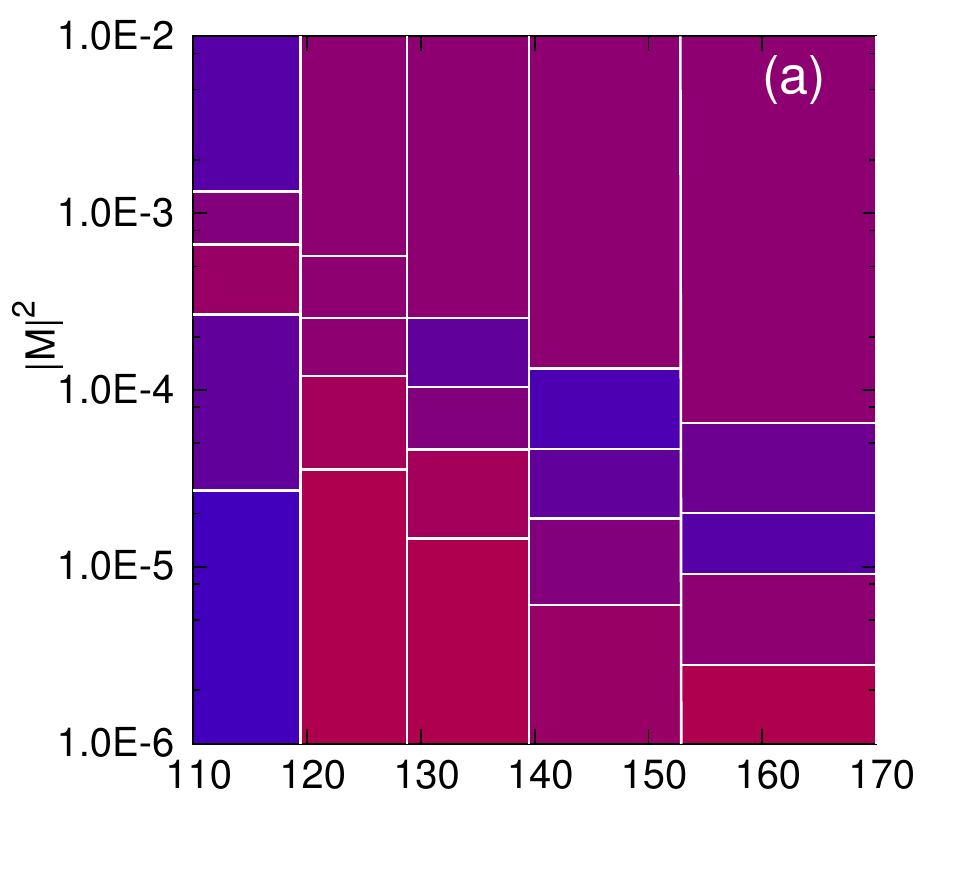} 
\includegraphics[height=0.27 \columnwidth]{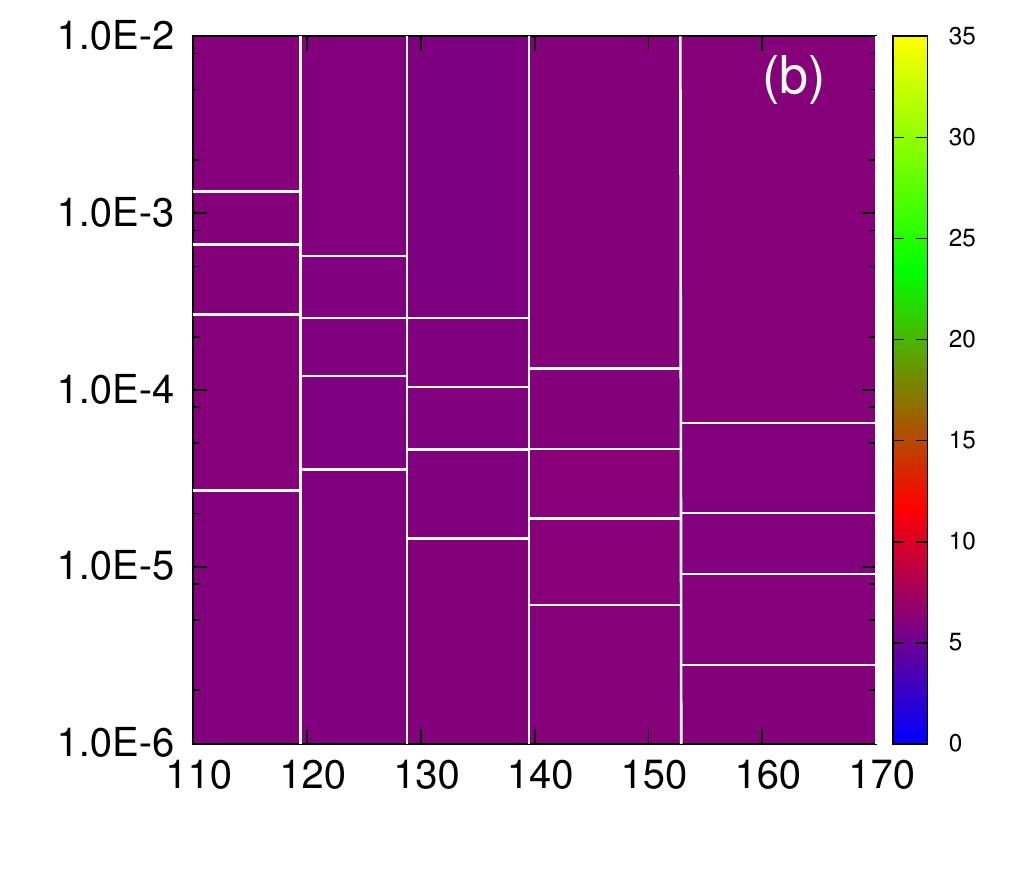}\\
\includegraphics[height=0.27 \columnwidth]{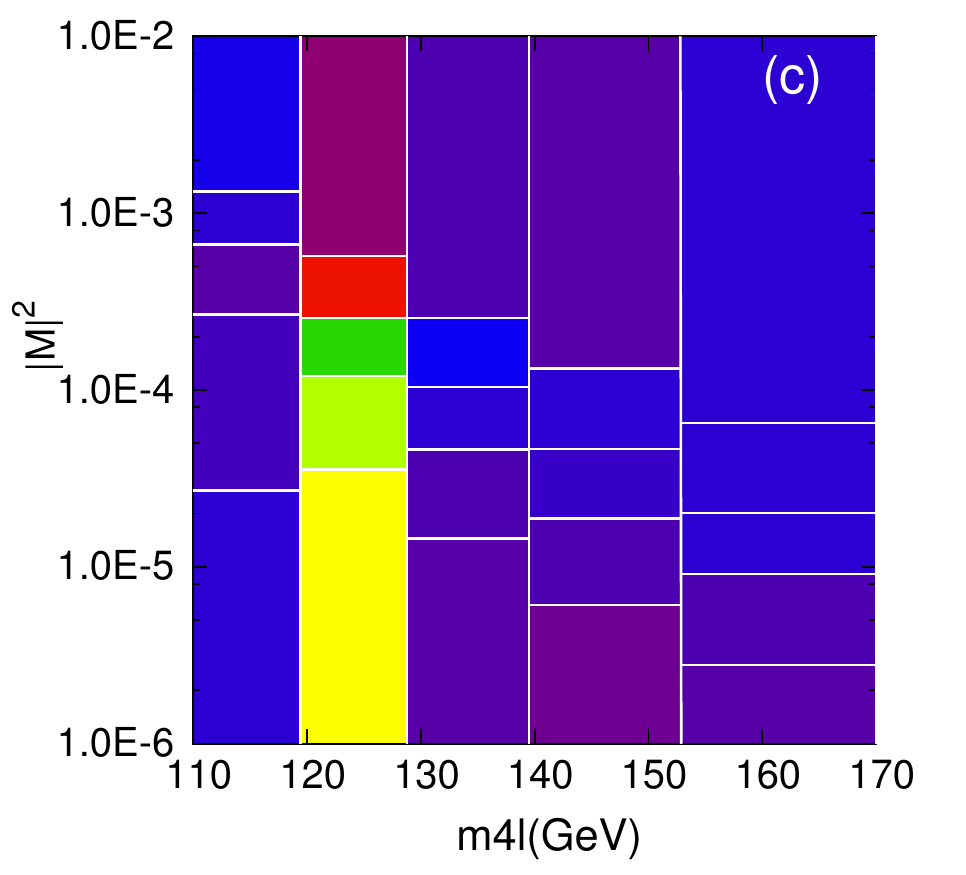} 
\includegraphics[height=0.27 \columnwidth]{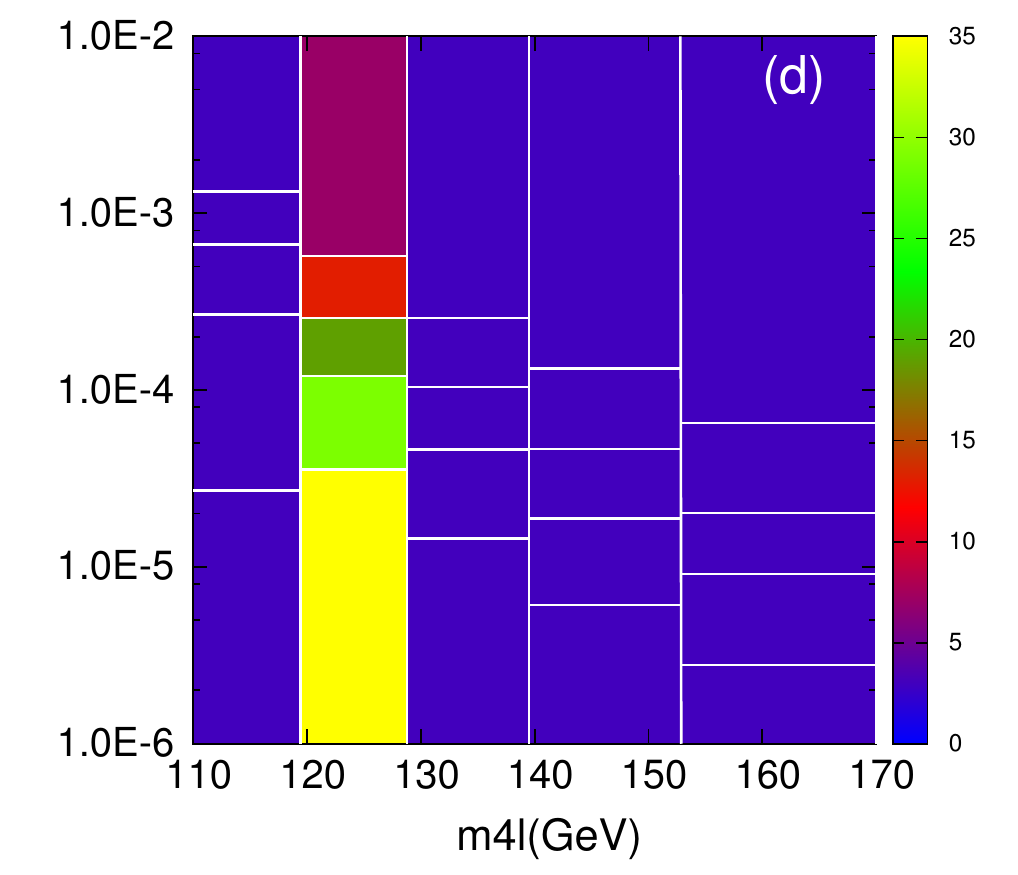}\\
\end{center}
\caption{\label{fig:quantile}
Quantile bins with respect to $m_{4\ell}$ (x-axis) and $|{\mathcal
  M}|^2_{bg}$ (y-axis).
These bins have equal cross sections for the background ($q\bar{q} \to
2e2\mu$) process.  
The figures above show the number of events in each bin
from $150$ background $q\bar{q} \to 2e2\mu$ events (panels in
the top row), or $75$ $125$-GeV Higgs signal and $75$ background events
(panels in the bottom row).  The left column panels are obtained from
one $150$ event pseudo-experiment, while the panels in the right
column are obtained from the average of $400$ such pseudo-experiments.}
\end{figure}

\begin{figure}[th]
\begin{center}
\includegraphics[width=0.3\columnwidth]{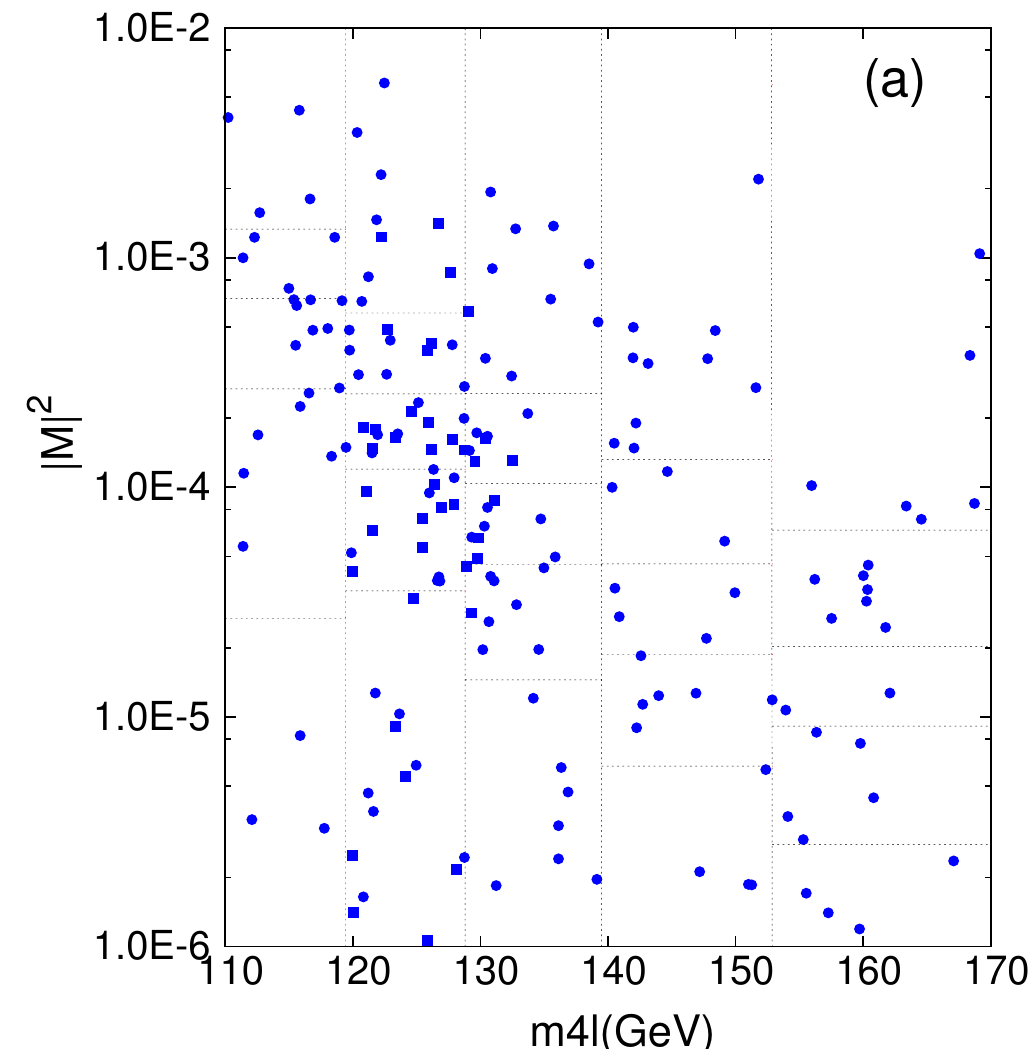} 
\includegraphics[width=0.3\columnwidth]{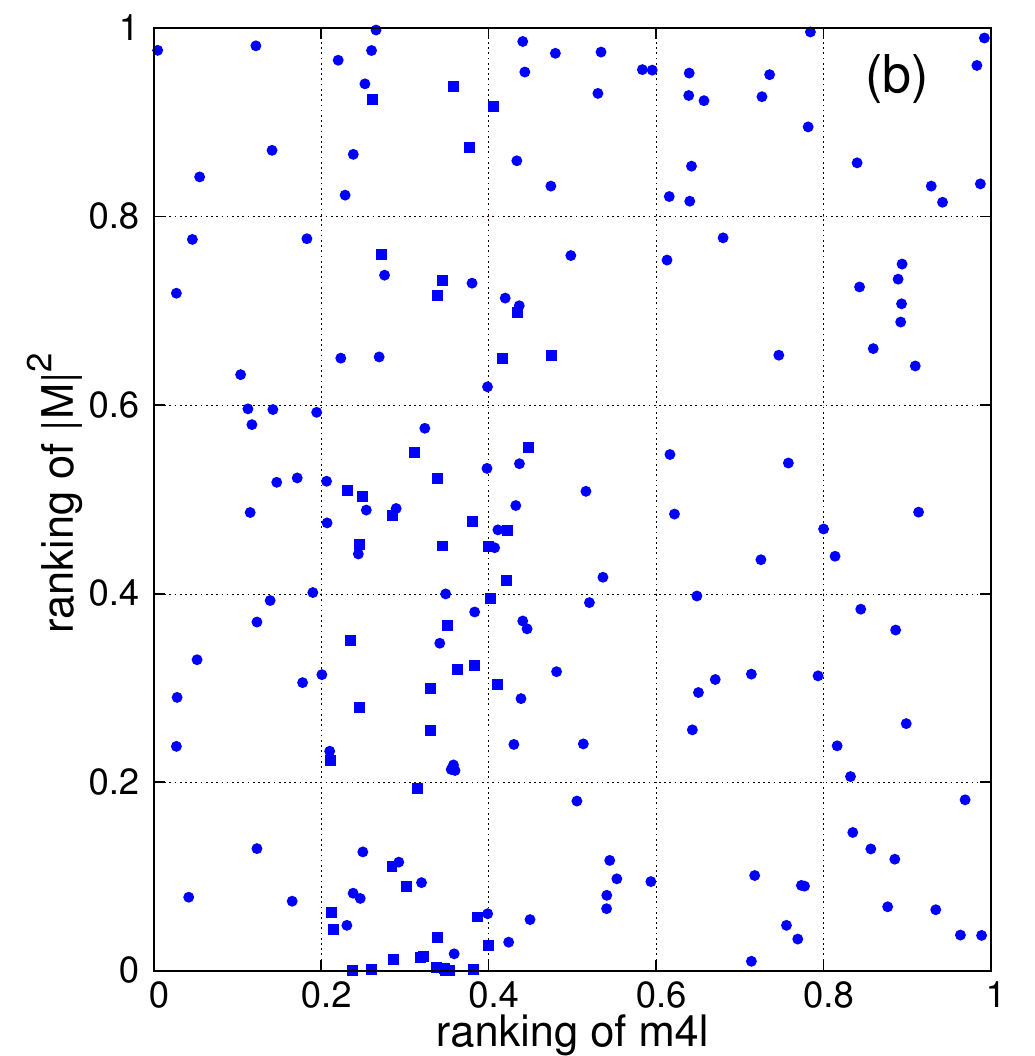} 
\end{center}
\caption{\label{fig:quantile-flattening}
The figures show $50$ 125-GeV Higgs $gg \to H \to 4\ell$
simulated events (squares) and $150$ $q\bar{q} \to 4\ell$ simulated
events (circles). 
The four-lepton invariant mass and pdf-weighted background squared matrix
element are plotted in panel (a); panel (b) utilizes the ranking
variable defined in equation~(\ref{eq: rank int}).
In each case, the quantile bin boundaries are represented by the grid
of dotted lines.
}
\end{figure}

We note that we can obtain quantile bins for $r(|{\mathcal
  M}|^2_{bg})$, possibly in tandem with other variables, directly\footnote{
Quantile bins have previously been employed in studies of the LHC
inverse problem~\cite{AKTW}.
}.
For a single variable, $\xi$, we find $n+1$ values $\eta_1, \eta_2,
..., \eta_{n+1}$ such that
\begin{equation}
\int_{\eta_i}^{\eta_{i+1}} \frac{dN}{d\xi} d\xi = 1/n.
\end{equation}
For several variables $\xi_i$, we also demand an equal number of
expected events in each bin, e.g., in two dimensions,
we choose values of $\xi_1$: $\eta_{1,1}$, $\eta_{1,2}$, ...,
$\eta_{1,n+1}$
and values of $\xi_2$: $\eta_{2,1}$, $\eta_{2,2}$, ..., $\eta_{2,n+1}$, such that
\begin{eqnarray}
\int_{\eta_{1,i}}^{\eta_{1,i+1}} \int_{\eta_{2,j}}^{\eta_{2,j+1}} \frac{d^2N}{d
  \xi_1 d \xi_2} d \xi_1 d \xi_2 = \frac{1}{n^2}.
\end{eqnarray}
Thus we can consider kinematic variables, such as invariant mass or
missing energy, in addition to a matrix-element-based variable.
We show toy examples of this approach in figures~\ref{fig:quantile}
and~\ref{fig:quantile-flattening}, which depict the distribution of
four-lepton events at the 8 TeV LHC.  The two variables used here are
the four-lepton invariant mass, $m_{4 \ell}$, and the background {\tt
  MEKD} value. 

Figure~\ref{fig:quantile}  shows the results of a simulated
experiment.  We have analyzed the data using quantile bins in
$m_{4\ell}$ and $|{\mathcal M}|^2_{bg}$, which were constructed using the
background hypothesis.
In each quantile bin, we plot either the number of events from $150$
background $q\bar{q} \to 2e2\mu$ events (top row), or $75$ $125$-GeV
Higgs signal and $75$ background events
(bottom row).  The left column panels were obtained from one $150$
event pseudo-experiment, while the panels on the right were obtained
from the average of $400$ such pseudo-experiments.
Figure~\ref{fig:quantile-flattening} also shows how flattening helps
the eye tell whether an excess is due to signal.  In this figure, the
ratio of signal to background events has been
changed from the realistic value 1:1 to the more challenging value of 1:3.  
Still, the presence of the new signal can be inferred, even by eye,
due to the advantages in visualization obtained from the use of
ranking variables.

\subsection{Flattening with Respect to All the Variables}
\label{subsec:3}
In principle, we can flatten the background distribution with
respect to a complete set of kinematic variables.  This can be done
numerically (though one would expect this to be numerically
challenging) or analytically, provided the analytic description of the
background is sufficiently accurate.

\begin{figure}[th]
\begin{center}
\includegraphics[width=0.45\columnwidth]{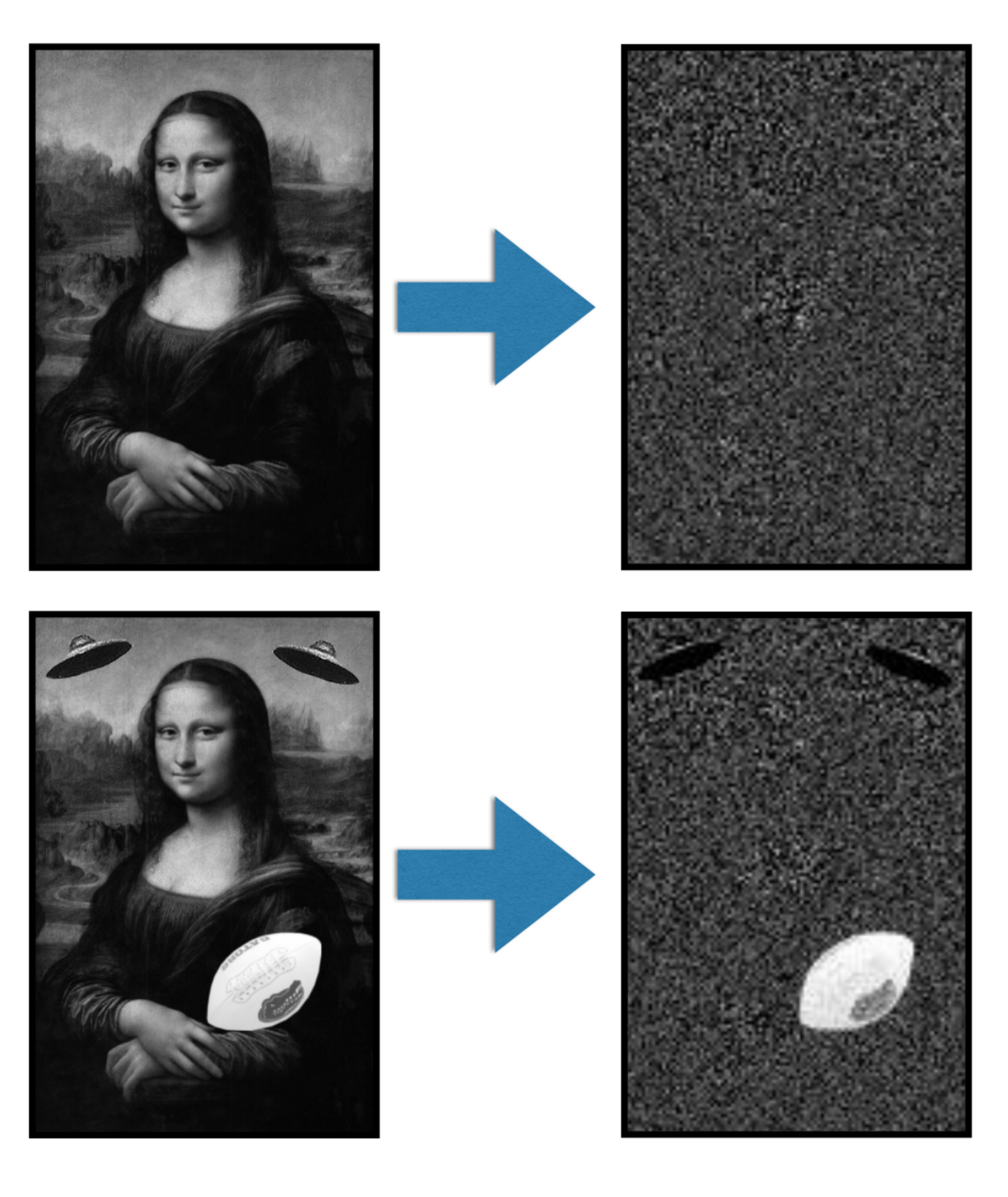}\\
\end{center}
\vspace{-15 pt}
\caption{\label{fig:mona}
The flattening procedure described in subsection~\ref{subsec:3}, as
applied to a histogram obtained only from the background (top row) or
when both signal and background are present (bottom row).  In these
figures the Mona Lisa plays the role of the background, while the
American football and flying saucers represent the signal.
}
\end{figure}

To be more precise, let the background (after detector simulation, etc.)
be described by the differential distribution $d^n N / d \bm{\xi}$.
Weighting each background event by
$1/(d^n N / d \bm{\xi})$, produces a distribution that is flat in the
full $n$-dimensional space of values.  
If data events are weighted according to this procedure, 
a signal will be visible as a deviation from flatness.  
We illustrate this procedure in figure~\ref{fig:mona}.


\section{Conclusions}

Variables based on the background squared matrix element can be used
to search for new physics signals at the LHC in a relatively
model-independent way.  
We look forward to the use of such techniques in future LHC analyses.

\Acknowledgements
J.G. and K.M. would like to thank their CMS colleagues for useful
discussions.  All of the authors would also like to thank D.~Kim and
T.~LeCompte for stimulating conversations and L. da Vinci for
assistance with figure~\ref{fig:mona}.
Work supported in part by U.S. Department of Energy Grant ER41990.

\end{document}